\documentstyle[ijp,twoside,multicol,epsf,color]{ijptex}

\def\prb{{\it Phys. Rev. B }}
\def\pra{{\it Phys. Rev. A }}
\def\prl{{\it Phys. Rev. Lett. }}
\def\ajp{{\it Am. J. Phys. }}
\def\pla{{\it Phys. Lett. A }}

\def\ibmjrd{{\it IBM J. Res. Dev. }}
\def\pjp{{\it Pramana J. Phys. }}
\begin{document}
\setcounter{page}{580}

\draft


\title{Loss of interference in an Aharonov-Bohm ring}

\author{Sandeep K. Joshi, Debendranath Sahoo\footnote{Permanent address: 
Materials Science Division,IGCAR,\\ Kalpakkam 603 102, Tamil Nadu, India.} 
and A. M. Jayannavar}

\address{Institute of Physics, Sachivalaya Marg, \\Bhubaneswar 751 005, 
India}

\email{joshi@iopb.res.in}


\maketitle

\begin{abstract}
We study a simple model of dephasing of Aharonov-Bohm oscillations in the
transmission of an electron across a mesoscopic ring. A magnetic impurity
in one of the arms of the ring couples to the electron spin via an
exchange interaction. This interaction leads to spin flip scattering and
induces dephasing via entanglement. This is akin to the models evoked
earlier to explain destruction of interference due to which-path
information in double-slit experiments. Total transmission is found to be
symmetric under flux reversal but not the spin polarization.
\end{abstract}



\thispagestyle{empty}
\pagestyle{myheadings}
\markboth{{\it S. K. Joshi, D. Sahoo and A. M. Jayannavar}}{
{\it Loss of interference in an Aharonov-Bohm ring}}

\begin{multicols}{2} 
\section{Introduction} 
The notion of intrinsic decoherence and dephasing of a particle
interacting with its environment is being pursued actively in the area of
mesoscopic physics. This is important from the basic point of view of
understanding the emergence of classical behavior from the quantum
dynamics.  In this area, study of transmission of electrons across a
mesoscopic Aharonov-Bohm ring occupies a prominent place from experimental
as well as theoretical viewpoint \cite{imry,datta,psd,webb_ap,gia,wgtr}.  
Generally in these systems, such a transition can be observed as a
function of temperature. At very low temperatures the inelastic scattering
length is much larger than the sample dimensions and as a result the
transport is completely phase coherent i.e., it is dominated by quantum
interference effects. At very high temperatures the inelastic scattering
length is much smaller than the sample dimensions which leads to Ohmic
transport or classical behavior. This process is also referred to as
dephasing owing to the loss of interference as a result of the
randomization of the interfering particle's phase.

In a double slit setup, interference results from the lack of knowledge of
(or indistinguishability of) the electron path. Thus a measurement of
which path the electron has taken, wipes out the interference pattern.  
It is known that in a ring interferometer the electron affects the
environment and changes its state differently in the two arms of the ring
thereby affecting the interference. This amounts to a measurement of the
path of the interfering particle by the environment resulting in loss of
interference. Such interferometers are thus also termed as ``which-path''
detectors. In an alternate picture, the environment affects the electron
phase differently in the two arms, thus randomizing their relative phase
difference leading to dephasing. The two views were shown to be
equivalent\cite{sai}.

It is well known that the electron-environment entanglement can also lead
to dephasing\cite{schulman}. However, unlike other approaches,
entanglement leads to dephasing in absence of any energy transfer
\cite{sai}. Thus motivated we consider a simple model of dephasing in
Aharonov-Bohm ring with a spin-half impurity (spin-flipper) in one
arm. This example also serves to illustrate the effect of multiple
reflections on "which-path" detection. We also show that the
spin-polarization which is related to the spin conductance is asymmetric
in flux reversal.

\begin{figure}
\protect\centerline{\epsfxsize=3.0in \epsfbox{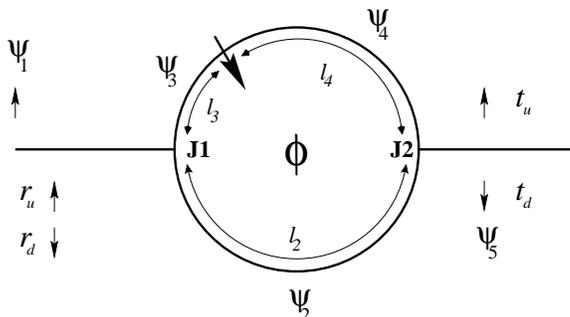}}
\caption{Mesoscopic ring with Aharonov-Bohm flux $\phi$ threading
  through the ring and a magnetic impurity in one arm of
  the ring.}
\label{ring}
\end{figure}

\begin{figure}
\protect\centerline{\epsfxsize=3.2in \epsfbox{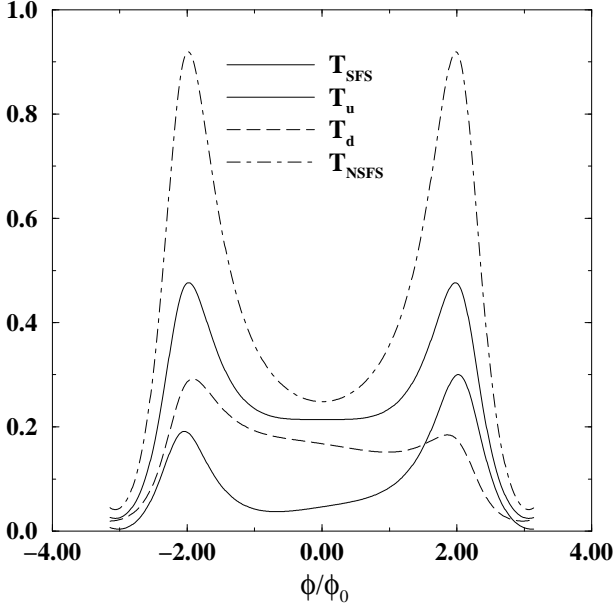}}
\caption{Plot of total transmission coefficient $T_{\mbox{{\tiny NSFS}}}$
for NSFS, total transmission coefficient $T_{\mbox{{\tiny SFS}}}$ for SFS,
and spin-up transmission coefficient $T_u$ and spin-down transmission
coefficient $T_d$ for the spin-flip scattering case. The interaction
strength $G=10.0$. }
\label{symm}
\end{figure}

\section{Model} 
Consider a spin-up electron incident from left onto the ring (see Fig.
\ref{ring}) in the presence of Aharonov-Bohm flux $\phi$. The electron
spin ($\vec{\sigma}$) is coupled to the spin of the flipper ($\vec{S}$)
via an exchange interaction $-J \vec{\sigma} \cdot \vec{S} \delta(x-l_3)$.
The vector potential along the ring circumference is $A=\phi/l$ where
$l=l_2+l_3+l_4$ is the ring perimeter, $l_2$ being the length of the lower
arm of the ring and $l_3$($l_4$) is the distance of the impurity from
junction J1(J2) in the upper arm.  The exchange interaction conserves the
total spin angular momentum and its $z$-component. This leads to two kind
of scattering processes depending upon the initial state of the impurity
spin namely, spin flip scattering (SFS) when the initial state of impurity
is down or no spin-flip scattering (NSFS) when it is up. It should be
noted here that none of these two processes involve any exchange of energy
and are perfectly elastic scattering events.  In case of NSFS the problem
at hand reduces to that of a simple potential scattering. However, the
exchange interaction does lead to entanglement of the electron and
impurity wavefunctions. By using the standard quantum waveguide
theory\cite{wgtr,ajp} and applying continuity of wavefunctions and current
conservation conditions at the impurity site and the junctions J1 and J2
we have calculated the probabilities of transmission of the electron as a
spin-up electron ( $T_u=|t_u|^2$ ) and spin-down electron ( $T_d=|t_d|^2$
). For details, we refer the reader to Ref. \onlinecite{xxx}. The total
transmission probability is simply the sum of the up and the down
transmission probabilities i.e., $T=T_u+T_d$ and the spin-polarization is
given by $\chi=(T_u-T_d)/T$.  The lengthy analytical expressions restrict
us to a graphical presentation of our results. In the following we have
set $\hbar=2m=1$, $kl=1$ and the value of the interaction strength
$G=2mJ/\hbar^2$ is given in dimensionless units. We have chosen
$l_2/l=0.5$, $l_3/l=0.15$, $l_4/l=0.35$ for the results presented below.

\begin{figure}
\protect\centerline{\epsfxsize=3.2in \epsfbox{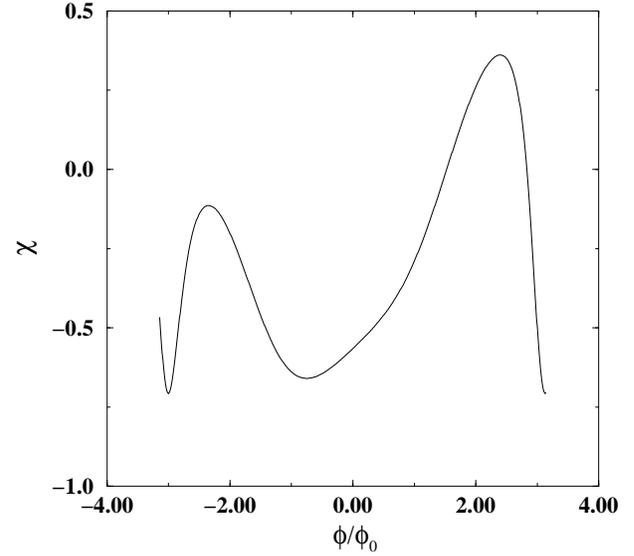}}
\caption{ Spin polarization ($\chi$) as a function of the flux $\phi$
  for interaction strength $G=10.0$. }
\label{spinpol}
\end{figure}

\section{Results and discussion}
In figure \ref{symm} we show the plot of total transmission coefficient
$T_{\mbox{{\tiny NSFS}}}$ for no spin-flip scattering case (where incident
electron spin is up $\sigma_z=1/2$ and initial impurity spin is down
$S_z=1/2$) and total transmission $T_{\mbox{{\tiny SFS}}}$, spin-up
transmission $T_u$, spin-down transmission $T_d$ coefficients for the
spin-flip scattering case (here incident electron spin is up
$\sigma_z=1/2$ and initial impurity spin is down $S_z=-1/2$) as a function
of $\phi/\phi_0$, $\phi_0=hc/e$ being the flux quantum. Perhaps the most
easily recognizable feature of the figure is the symmetry of both
$T_{\mbox{{\tiny NSFS}}}$ and $T_{\mbox{{\tiny SFS}}}$ under flux reversal
and the $2\pi\phi_0$ flux periodicity of the AB oscillations. Such
expected periodic oscillations in the transmittance have been observed
experimentally\cite{webb_ap}. However, individually the spin-up and
spin-down transmission coefficients, although having the same $2\pi\phi_0$
flux periodicity, are not symmetric under flux reversal.  The problem of
transport in presence of spin-flip scattering, in spite of absence of any
inelastic scattering reduces to the multichannel case. The symmetry
properties noted above are consistent with reciprocity relations for
transport in multichannel systems and are a consequence of the general
symmetries of the Hamiltonian\cite{butt_ibm}. This asymmetry in the
individual spin-up and spin-down components of transmission presents
itself in the asymmetry observed in the spin-polarization $\chi$ as seen
in Fig. \ref{spinpol}. The spin-conductance in spin-polarized transport is
related to the spin-polarization\cite{sds} and therefore is asymmetric
under the flux reversal. The zero temperature total electrical conductance
is a sum of total transmission coefficients for all the four possible
initial conditions ($\sigma_z$,$S_z$)=($\pm$1/2,$\mp$1/2) and is symmetric
under the flux reversal.

\begin{figure}
\protect\centerline{\epsfxsize=3.2in \epsfbox{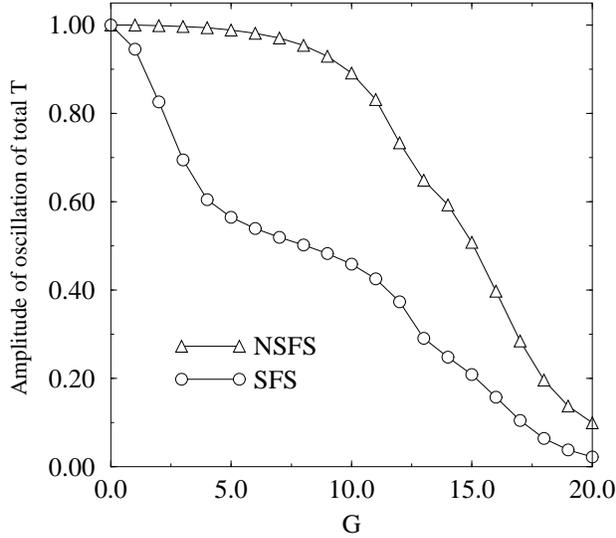}}
\caption{ Variation of amplitude of oscillation of total transmission  
coefficient with the interaction strength $G$ for the two cases of
spin-flip scattering (SFS) and no spin-flip scattering (NSFS).}
\label{amp}
\end{figure}

The second important feature which Fig. \ref{symm} exhibits and was
mentioned above but not emphasized is that $T_d$ also shows an
interference pattern (AB oscillations with a flux periodicity of
$2\pi\phi_0$) in the SFS case. This seems to contradict the naive
expectation that a spin-flip would amount to path detection and therefore
one should not, in principle, observe any interference pattern for
spin-down component of transmission. Realizing that the above expectation
rests on the belief that only two forward propagating partial waves, one
in each arm of the ring, produce the interference pattern, helps to
clarify the situation. In the present geometry there are infinitely many
partial waves, owing their existence to the multiple reflections induced
by the reflection and transmission at the junctions J1, J2 and impurity
site, which superimpose to produce the interference pattern. To clarify
the point further consider just one possible path that the electron could
take out of the infinitely many. The incident spin-up electron moving in
the upper arm of the ring gets spin-flipped and reflected at the impurity
and finally traverses the lower arm. This partial wave will then interfere
with the spin-flipped component transmitted across the impurity in the
upper arm to give rise to an interference pattern for $T_d$. Thus the
multiple reflections erase the "which-path" information. Naturally, then
the question arises - will we still observe dephasing in such a situation.
Figure \ref{amp} answers the question in the affirmative. The signature of
dephasing is that the amplitude of AB oscillations of total transmission
coefficient (or visibility factor) for the SFS case is always smaller than
that for the NSFS case for all non-zero values of interaction strength
$G$. For $G=10.0$, by comparing the $T_{\mbox{{\tiny SFS}}}$ with the
$T_{\mbox{{\tiny NSFS}}}$ reduction of amplitude can be seen explicitly
from Fig. \ref{symm}.  The reduction in the amplitude of oscillations of
the SFS case as compared to the NSFS case indicates dephasing. Thus we see
that the spin-flipper acts as a dephasor.

\section{Summary}
In summary, we have studied the electron transmission across a AB-ring
geometry with a spin half impurity in one arm of the ring using the
quantum waveguide theory.  The electron interacts with the impurity via an
exchange interaction. The naive expectation of vanishing of interference
pattern for the spin-down transmission due to which-path detection is to
be modified in the light of the important role played by multiple
reflections. The reduction in the amplitude of oscillations of the total
transmission coefficient for SFS in comparison to that for NSFS, clearly
brings out the feature of dephasing in this simple model.  Moreover, it is
important to note that the dephasing in this model is in the absence of
any inelastic scattering.  The study also reveals the asymmetric nature of
the spin-polarized transport as against the symmetric two probe
conductance.  Our further studies\cite{colin} have shown that such a
dephasor is not able to suppress the other well known quantum effect
namely, the current magnification\cite{psd,wgtr}. We believe that this
effect will be suppressed only in the presence of inelastic scattering.
Further work in regard to transport properties and additional resonances
due to spin-flip is in progress.

\label{lastpage}

\end{multicols}

\begin{references}

\bibitem{imry} Y. Imry, {\it Introduction to Mesoscopic Physics}
  (Oxford University, New York, 1997).

\bibitem{datta} S. Datta, {\it Electronic transport in mesoscopic systems}
(Cambridge University Press, Cambridge, 1995).

\bibitem{psd} P. S. Deo and A. M. Jayannavar, \pjp , (2000) in print, 
cond-mat/0006035.

\bibitem{webb_ap} S. Washburn and R. A. Webb, {\it Adv. Phys.} {\bf 35}, 375 
(1986).

\bibitem{gia}Y. Gefen, Y. Imry, and M. Ya. Azbel, \prl {\bf 52}, 129 (1984).

\bibitem{wgtr} A. M. Jayannavar and P. S. Deo, \prb {\bf
    49}, 13 685 (1994); {\bf 51}, 10 175 (1995); P. S. Deo and A. M.
  Jayannavar, \prb {\bf 50}, 11629 (1994); T. P. Pareek, P. S. Deo and
  A. M.  Jayannavar, \prb {\bf 52}, 14 657 (1995); and references therein.

\bibitem{sai} A. Stern, Y. Aharonov and Y. Imry, \pra
    {\bf 41}, 3436 (1990).


\bibitem{schulman} L. S. Schulman, \pla {\bf 211}, 75 (1996).






\bibitem{ajp} O. L. T. de Menezes and J. S. Helman, \ajp
    {\bf 53}, 1100 (1985).

\bibitem{xxx} S. K. Joshi, D. Sahoo and A. M. Jayannavar, preprint
  cond-mat/0007414

\bibitem{sds} S. Das Sarma, J. Fabian, X. Hu, I. Zutic,
  Superlattices and Microstructures {\bf 27}, 289 (2000); I. Zutic and
  S. Das Sarma, \prb {\bf 60}, 16 322 (1999).

\bibitem{butt_ibm} M. B\"{u}ttiker, \ibmjrd {\bf 32},
  317 (1988).

\bibitem{colin} Colin Benjamin, S. K. Joshi, D. Sahoo and A. M. Jayannavar, 
unpublished

\end{references}
\end{document}